\documentstyle{mn}
\input epsf.sty

\def\Rg{R_{\rm g}}

\def\Rms{R_{\rm ms}}
\def\NH{N_{\rm H}}

\def\Rin{R_{\rm in}}
\def\Din{D_{\rm in}}
\def\Rout{R_{\rm out}}
\def\Ka{{\rm K}\alpha}

\def\ginga{{\it Ginga}\ }

\def\MEdd{\dot M_{\rm Edd}}
\def\LEdd{L_{\rm Edd}}

\def\Teff{T_{\rm eff}}
\def\Tcol{T_{\rm col}}

\def\mdot{\dot m}
\def\rtr{r_{\rm tr}}
\def\LX{L_{\rm X}}

\def\Firr{F_{\rm irr}}
\def\relrepr{{\tt relrepr}\ }
\def\MSun{M_{\odot}}
\def\gs{{GS~2023+338}\ }
\def\tes{{\tau_{\rm es}}}

\def\Or{\Omega_{\rm r}}
\def\On{\Omega_{\rm r, non-smeared}}

\title[Spectral evolution of GS 2023+338]
{X--ray spectral evolution of GS~2023+338 (V404 Cyg) during 
decline after outburst}
\author[P.T. \.{Z}ycki, C. Done and D.A. Smith]
     {Piotr T. \.{Z}ycki,$^{1,3}$ Chris Done$^1$ and 
     David A. Smith$^{2}$\thanks{Present address: NASA/Goddard Space 
     Flight Center, Greenbelt, MD 20771, USA}\\ 
        $^1$ Department of Physics, University of Durham,
       South Road, Durham DH1 3LE; chris.done@durham.ac.uk \\
        $^2$ Department of Physics and Astronomy, University of Leicester,
            University Road, Leicester LE1 7RH \\
        $^3$ Nicolaus Copernicus Astronomical Center, Bartycka 18, 00-716 
            Warsaw, Poland; ptz@camk.edu.pl}
 
\date{6 November 1998}
 
\pagerange{\pageref{firstpage}--\pageref{lastpage}}
\pubyear{1999}

\begin{document}

\maketitle
 
\begin{abstract}

We have re-analyzed archival {\it Ginga}\/ data of the soft X--ray transient
source GS~2023+308 (V404 Cyg) covering the decline phase of its May 1989
outburst. Our spectral modeling includes the relativistically smeared Compton
reflected continuum and iron $\Ka$ fluorescent line near 6.5 keV produced by
X--ray illumination of the accretion disc.  This gives a powerful diagnostic
of the accretion geometry, with the amplitude of the reprocessed components
showing the solid angle subtended by the disc, while the detailed shape of
the relativistic smearing shows how close this material is to the black hole
event horizon. 

We find that reflection is always significantly present in the spectra, but
that its fractional contribution decreases as the decline progresses. The
amount of relativistic smearing is also consistent with decreasing during the
decline, although this is poorly constrained except for the brightest
spectra.  One plausible scenario explaining this evolution is of an optically
thick disc with inner radius increasing as a function of time, with the X--ray
source in the form of a central corona. This is similar to the evolution
inferred for other X--ray transient sources, such as Nova Muscae, except that
the underlying power law spectrum of \gs stayed constant as the disc geometry
changed.  This challenges the underlying assumption of almost all models for
the spectra of accreting black holes, namely that the hard X--rays are formed
by Comptonization of seed photons from the accretion disc. 

\end{abstract}
 

\label{firstpage}
  
\section{Introduction}

The 'no hair' theorem for black holes states that they can be described solely
in terms of their mass and spin (and charge, although this is unlikely to be
important in physically realistic situations). Black holes are then completely
characterised by two parameters, making them much less complex than e.g.\  
neutron 
stars, where there can be a strong magnetic field, and where there is an
uncertain equation of state for the dense material. Black hole binary systems
are then the simplest objects in which to study the physics of accretion flows
in strong gravity, and have the further advantage that the orbital parameters
are often well studied so that the inclination, mass and distance of the system
are tightly constrained. 

Many of the black hole binary systems show dramatic outbursts where the
luminosity rises rapidly from a very faint quiescent state to one which 
is close
to the Eddington limit, $L_{\rm Edd}$, and then declines approximately
exponentially over a period of months. These systems are the soft X--ray
transients (hereafter SXT; see Tanaka \& Lewin 1995, Tanaka \& Shibazaki 1996
for reviews), with the outburst triggered by the classic disc
instability which occurs as Hydrogen starts to become partially ionised 
(Meyer \& Meyer-Hofmeister 1981; Smak 1982; see Osaki 1996 for review).
The dramatic increase in mass accretion rate onto the central object gives rise
to strong X--ray emission. This illuminates the disc, and controls its 
evolution
during the decline phase (King \& Ritter 1998, King 1998), and in quiescence
(King, Kolb \& Szuszkiewicz 1997).

The SXT also show spectral evolution during the outburst and decline. 
In general, when the luminosity is close to the Eddington limit, 
their X--ray spectra show a strong thermal component at $\sim 1$ keV, 
accompanied by a
strongly variable power law (very high or flare state). As the luminosity
decreases then the power law component decreases in importance, giving the high
or soft state which is dominated by the thermal component. There is then a
marked transition at $L \sim 0.1 L_{\rm Edd}$ as the thermal component 
decreases
in both luminosity and temperature, and a hard, strongly variable power law
dominates the X--ray spectrum (the low or hard state).  This spectrum remains
fairly stable to very low luminosities, where the source is in the quiescent or
off state (Miyamoto et al.\ 1992; Nowak 1995; Tanaka \& Lewin 1995; 
Tanaka \& Shibazaki 1996).

According to current paradigm the thermal component originates in an optically
thick accretion disc for which the Shakura--Sunyaev (1973; hereafter SS)
solution is commonly used, while the hard power law is produced by hot,
optically thin plasma. The geometry inferred for the low/hard state from
modelling the broad band (1--200 keV) spectrum of both transient and persistent
sources (Dove et al.\ 1997; Gierli\'{n}ski et al.\ 1997a; Poutanen, Krolik \&
Ryde 1997; see Poutanen 1998 for a review) is that of a geometrically thin, 
cool
accretion disc which truncates at some transition radius, $\rtr$, which is
rather larger than the innermost stable orbit at $6 \Rg$ ($\Rg\equiv G M/c^2$),
with the inner region filled 
by a quasi-spherical, X--ray hot plasma.

Another independent diagnostic of the accretion geometry is produced as the
X--rays illuminate the disc. 
An iron fluorescence line and reflected continuum are formed wherever
hard X--rays illuminate optically thick material (George \& Fabian 1991; Matt,
Perola \& Piro 1991), but around a black hole these
spectral features should be strongly smeared by the combination of Doppler
effects from the high orbital velocities and strong gravity (Fabian et al.,
1989). Modelling of the detailed shape of this relativistic smearing directly
constrains the distance of the cool reflecting material from the black hole,
while the amount of reflection and line constrains the geometry.  Results from
this approach from low state spectra of both transient and persistent Galactic
Black Hole Binaries are consistent with the geometry described above (Ueda, 
Ebisawa \& Done 1994; Ebisawa et al.\ 1996; \.{Z}ycki, Done \& Smith 1997,
hereafter Paper I; \.{Z}ycki, Done \& Smith 1998, hereafter ZDS98; Done \&
\.{Z}ycki 1998).

The inferred high state geometry is rather different. Both broad band spectra
and detailed modelling of the reprocessed component give a consistent picture 
in which the optically thick disc extends all the way down to the last 
stable orbit (Poutanen et al.\ 1997; Gierli\'{n}ski et al.\ 1997b; ZDS98; 
Cui et al.\ 1997). The transition to the high state then seems to involves 
a decrease of
$\rtr$ and consequently an increase of the disc thermal emission at the expense
of the power law component. Further evolution to the very high state 
presumably 
requires a redistribution of the energy generation so that the relative 
contribution of a coronal source increases.

There is no current model able to compellingly explain all this evolution.
Perhaps the best to date is that of Esin, McClintock \& Narayan (1997), who go
some way towards constructing the global dynamics as a function of mass
accretion rate. At low mass accretion rates there are least two thermally and
viscously stable solutions of the accretion flow. One is the familiar optically
thick, geometrically thin, SS disc, where the energy released is efficiently
radiated via blackbody cooling. Another is a hot, optically thin, geometrically
thick flow in which the radiative cooling is rather inefficient so that radial
energy transport (advection) is important (advection dominated accretion flows,
hereafter ADAF; see e.g.\ the review by Narayan  1997).
For the hot solution, an increasing mass accretion rate, $\mdot$
means that the flow density and, consequently, its
radiative efficiency increases. Eventually, at some critical accretion rate
$\mdot=\dot m_{\rm crit}$, the cooling becomes efficient enough to 
collapse the hot solution down into an SS disc. 
Esin et al.\ (1997) calculate that this transition occurs
for luminosities $\sim 0.08 L_{\rm Edd}$, so they identify the low--high state
change with the infalling material switching from an ADAF  to SS disc.

At least three of currently known transient black hole systems did not show
this canonical
behaviour.  GS~2023+338 (V404 Cyg), GRO~J0422+32 (Nova Persei 1992) and
GRS~1716-249 (Nova Oph 1993) remained
dominated by the non-thermal component throughout the entire decline phase.
For GRO~J0422+32 and GRS~1716-29 this is plausibly because the system 
luminosities were never
higher than $\sim 0.1 L_{\rm Edd}$ so that it never went above the low state
(Nowak 1995; Esin et al.\ 1998, Revnivtsev et al.\ 1998), but this is 
clearly not the case for GS~2023+338, where the luminosity reached almost
the Eddington limit (Tanaka \& Lewin 1995). This system is especially 
interesting since it was used
as an example for the ADAF-based models of SXT's in quiescence (Narayan,
Barret \& McClintock 1997), yet its outburst behaviour is apparently very
different from that predicted by the extension of this model to 
the overall evolution of SXT (Esin et al.\ 1997).

In this paper we present results of detailed spectral analysis of data 
obtained during the 1989 outburst of GS~2023+338 (Kitamoto et al.\ 1989), and 
use the relativistically smeared reprocessed component to constrain the 
accretion  geometry. The outburst was well
covered by \ginga from its first detection by All Sky Monitor on May 22 until
November 1989 (Tanaka \& Lewin 1995).  In this paper we concentrate on the
decline phase of the outburst after 1 June 1989, leaving detailed discussion of
the most dramatic ``day in the life'' of \gs (30 May) to a forthcoming paper.
For our analysis we have selected several data sets when the effect of
photo-electric absorption (another unusual feature of the source) was
small and we aim at describing the evolution of the accretion flow geometry in
the vicinity of the central black hole.  We then compare this evolution to
more typical cases of STX e.g.\ Nova Muscae 1991, and with the predictions of
the Esin et al.\ (1997) model.

\begin{figure*}
%gs2023/fig_cc.sm: fig_time
  \epsfxsize = 0.95\textwidth
  \epsfbox[10 490 600 680]{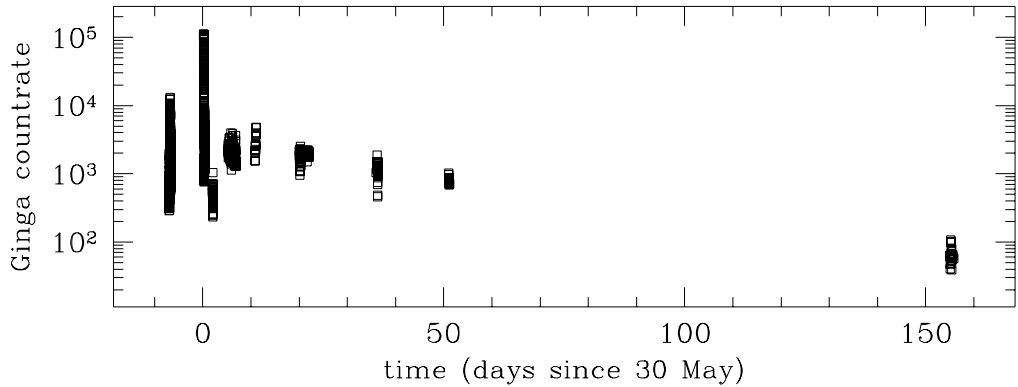}
\caption{\ginga light curve (1--20 keV) of \gs since the first pointed
 observations on 23 May 1989 until November 1989. The count rate was
 corrected for background, deadtime and aspect.  Dramatic flux and spectral
 variability
 was observed on 30 May. Flux variability occurred during each observation
 but the degree of spectral variability was changing with time. The upper
 envelope of the light curve (the 23.\ and 30.\ May points excepted) 
 follows an approximately exponential decline with e-folding time 
 $\sim 40$ days.
\label{fig:lcurve}}
\end{figure*}

\section{Data reduction and background subtraction}
 
The data were extracted from the original First Reduction Files in the usual 
manner using
the {\it Ginga\/} reduction software at Leicester University.  However,
unlike most X--ray sources, the spectrum of GS~2023+338 contributes a
significant number of counts above 24~keV.  This count rate is used as a
background monitor (Surplus above Upper Discriminator, or SUD), but here
it is contaminated by the source. The SUD rate minus the contribution from
the source was estimated from off-source data taken before and after the
source observation. We are able to do this because the satellite position
in orbit, and thus the intrinsic background, depends on two parameters
(excluding the phase of the 37-day orbital period of {\it Ginga\/}): the
angle from ascending node of the orbit and the longitude of the ascending
node (K.~Hayashida, private communication). We mapped the SUD rate from the
uncontaminated observations against both orbital parameters (dividing the
data into 20 degree bins).  For each data point from the observation of
the source we calculated the orbital parameters and replaced the contaminated
SUD value with that from the corresponding orbital parameters on the
uncontaminated grid.  We can then perform either a ``local'' or ``universal''
background subtraction as normal (Hayashida et al.\ 1989).
We chose the universal background since there were no background 
observations suitable for the local method.
 
To test the method we used the off--source (uncontaminated) data using
both the original and modified SUD values. The r.m.s.\ deviations of the
resulting spectra were less than 5--7 per cent over the 2--20 keV range. 
Secondly, we
perform a similar analysis on MCG-2-58-22 (Nov 1989 observation), comparing 
the ``universal'' background subtraction before and after inserting modified
SUD values. The results from the real and modified SUD data were in
agreement within the error bars, showing that the method is reliable. 
 
\begin{figure}
  \epsfxsize = 7.5 cm
  \epsfbox[10 170 550 700]{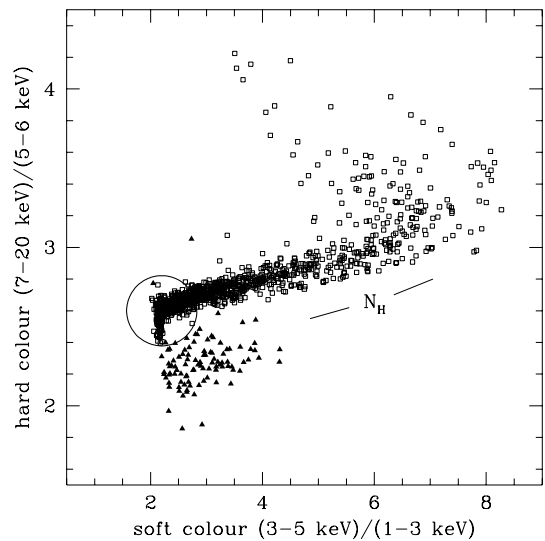}
\caption{
The colour--colour diagram of GS~2023+338 using the decline phase data i.e.\ 
after 1 June 1989. The 
well defined (almost) horizontal track is due to photo-electric absorption
with $\NH$ increasing to the right.
The turnoff point corresponds to $\NH\sim 4\times 10^{23}\ {\rm cm^{-2}}$,
when the spectrum above 5 keV begins to be affected. The circle marks a region
where the June -- July data for detailed spectral analysis were extracted from.
The November data are those marked with filled triangles
(see Table~\ref{tab:obslog}). \label{fig:colcol}}
\end{figure}

\section{Data selection for spectral analysis}
 
% 30/05/89 - 89.150

\begin{table*}
 \caption{Observations log for analysis of un-absorbed, decline phase spectra.
\label{tab:obslog}}
   \begin{tabular}{clcrcr}
label & data set & start date/time & end date/time & 
          start date$ - $ &  exposure (s) \\
      &          &                 &               &      
                      30 May (d) &          \\
\hline
1. & June 3.  &  89/154 ~~3/06 16:36:31 &  3/06 16:43:27 &  ~~4 & 416 \\
2. & June 4.  &  89/155 ~~4/06 12:11:59 &  4/06 12:20:07 &  ~~5 & 488 \\
3. & June 10. &  89/161  10/06 01:58:00 & 10/06 02:15:04 &  ~11 & 1024 \\ 
4. & June 19. &  89/170  19/06 06:29:40 & 19/06 07:46:40 &  ~20 & 1900 \\
5. & June 20. &  89/171  20/06 14:05:00 & 21/06 04:00:00 &  ~21 & 12180 \\
6. & July 6.  &  89/187 ~~6/07 09:47:00 &  6/07 10:10:28 &  ~37 & 1408 \\
7. & July 20. &  89/201  20/07 01:00:00 & 22/07 05:13:00 &  ~51 & 7650 \\
8. & November &  89/305 ~~1/11 00:20:00 &  1/11 22:00:00 &  155 & 19064 \\
\hline
   \end{tabular}
\end{table*}

The variability characteristics of these data were examined in detail by
Oosterbroek et al.\ (1996, 1997). Here we show the light--curves
(Fig.~\ref{fig:lcurve}) and colour--colour diagram (Fig.~\ref{fig:colcol})
for completeness.  There is remarkable variability both in flux and in
spectral shape (see also Tanaka \& Lewin 1995). 
Oosterbroek et al.\ (1997) showed that the well defined track 
(marked $N_{\rm H}$ on Fig.~\ref{fig:colcol}) is consistent with
being due to increasing photo--electric absorption. The left hand side of 
this track seems to end at a well defined point, plausibly
because this represents the unabsorbed primary spectrum. We extracted
datasets from this point in order to try to constrain the underlying
spectral shape since the beginning of June to the
last \ginga observation in 1989 (November), as detailed in
Table~\ref{tab:obslog}. 
 
\begin{table*}
 \caption{Results of model fitting to the sequence of decline phase spectra.
\label{tab:declspec}}
   \begin{tabular}{cccccclllr}
label & $\NH$ & 
  $k T_{\rm soft}$ (keV) & $\Din\ (\Rg)$ & 
  $\Gamma$  & $N$ & ~~~~ $\Or$ ~~~~ & ~~ $\xi$ ~~ & $\Rin\ (\Rg)$ & 
        $\chi^2/$dof \\
\hline
1. & $1.73\pm 0.22$ &
           $0.22^{+0.06}_{-0.09}$   &  $11^{+\infty}_{-9}$ &
           $1.67\pm 0.03$  & $1.42^{+0.09}_{-0.07}$ & $0.70^{+0.12}_{-0.10}$ &
           $0.1^{+2}_{-0.1}$    & ~$20^{+20}_{-9}$ &
           18.6/23 \\
2. &  $1.54^{+0.26}_{-0.20}$ &
          $0.24^{+0.07}_{-0.10}$   & $7^{+82}_{-4}$ &
          $1.70\pm 0.03$ & $1.21^{+0.07}_{-0.06}$  & $0.83^{+0.13}_{-0.11}$ &
          $0.1^{+2}_{-0.1}$     & ~$30^{+60}_{-11}$ & 
           21.6/23 \\
3. & $1.75^{+0.22}_{-0.24}$ &
           $0.27^{+0.03}_{-0.04}$   &  $8^{+5}_{-3}$   &
           $1.76\pm 0.03$ & $2.35^{+0.10}_{-0.13}$ & $0.88^{+0.12}_{-0.10}$ &
           $0^{+0.1}$        & ~$17^{+9}_{-5}$   &
           21.3/23 \\
4. & $2.04^{+0.22}_{-0.26}$ &
           $0.295\pm 0.010$        & $5.4^{+1.2}_{-1.1}$ &
           $1.72 \pm 0.03$ & $1.16\pm 0.06$  & $0.66^{+0.11}_{-0.07}$ & 
           $ 0^{+1} $        & ~$18^{+13}_{-7}$  &
           23.1/23 \\
5. & $1.28^{+0.30} _{-0.34}$ &
           $0.37^{+0.12}_{-0.04}$  & $1.6^{+1.0}_{-1.3} $ &
           $1.67\pm 0.03$ & $0.93^{+0.05}_{-0.06}$  & $0.50\pm 0.08$    &
           $1^{+8}_{-0.9} $   & ~$ 36^{+80}_{-17}$  &
           7.85/23 \\
6. & $2.05^{+0.22}_{-0.33}$ & 
           $0.23^{+0.04}_{-0.07}$   & $9.5^{+32}_{-4} $ & 
           $1.64\pm 0.03$ & $0.67\pm 0.04$  & $0.47\pm 0.09$  & 
           $1^{+9}_{-1}$   & ~$80^{+\infty}_{-55}$ &
           22.8/23 \\
7. & $1.19^{+0.36}_{-0.25}$ &
           $0.43^{+0.22}_{-0.07}$ & $0.6^{+0.7}_{-0.5}$ &
           $1.70\pm 0.03$ & $0.39\pm 0.03$ & $0.37\pm 0.07$  &
	   $12^{+30}_{-8}$ & ~$60^{+\infty}_{-35}$ &
	   19.9/23 \\
8. & $3.3\pm 0.6$ & 
           $0.30^{+0.04}_{-0.15}$ & $1^{+8}_{-0.4} $ &
           $1.93\pm 0.05$ & $(5.4^{+0.6}_{-0.4})\times 10^{-2}$ & 
                 $0.36^{+0.16}_{-0.10}$ &
           $60^{+90}_{-45}$ & $600^{+\infty}_{-565}$ &
	   26.3/22 \\
\hline
   \end{tabular}

\medskip
  $\NH$ -- hydrogen column of uniform absorber in units of 
    $10^{22}\,{\rm cm^{-2}}$ \\
  $D_{\rm in}$ -- inner radius of emitting disc as found from the
  amplitude of the soft thermal component. \\
  $\Gamma$ -- photon index of the primary power law. \\
  $N$ -- normalization of the primary power law (photons/cm$^2$/s/keV) \\
  $\Or$ -- amplitude of the smeared, reprocessed component. \\
  $\xi$  -- ionization parameter, $\LX/n r^2$. \\
  $\Rin$ -- inner radius of the disc as found from the magnitude
  of the relativistic effects.
\end{table*}

\section{Models}

We use a variety of models for the data analysis. The soft component is
described by a blackbody or a disc blackbody (i.e.\ multi-temperature
composition of blackbody spectra; Frank, King \& Raine 1992; Mitsuda et al.\
1984). For the hard component we use a power law, and its reflected continuum 
is modelled using the angle-dependent Green's functions of Magdziarz \& 
Zdziarski
(1995) as implemented in the {\tt pexriv} model in {\sc XSPEC} spectral 
analysis
package (Arnaud 1996). This calculates the ionization state of the material by
balancing photo--electric ionization from the irradiating hard power law 
against
radiative recombination, as in Done et al.\ (1992). The photo--electric
absorption edge energies for iron were corrected from the rather approximate
values given by Reilman \& Manson (1979) to those of Kaastra \& Mewe (1993).
The recombination rate is a function of temperature, while the 
photo--ionization
rate depends on the ionization parameter $\xi=\LX/n r^2$ where $\LX$ is the
ionising luminosity, $n$ is the density of the reflector and $r$ is the 
distance
of the reflector from the X--ray source. At a given temperature (which is 
chosen
rather than calculated self--consistently in this code) then the ion 
populations
are determined by $\xi$, with a rather weak dependence on the spectral shape
(see Done et al.\ 1992).
 
The iron $\Ka$ line corresponding to given ionization state, 
plasma temperature and spectral shape  is computed self consistently in our
code, rather than modelled as a separate component. We use the
Monte Carlo code of \.{Z}ycki \& Czerny (1994) to compute the line intensity
and profile (which is broadened by Compton scattering in the reflector and 
a possible range of ionization stages of iron). The line
is then added to the reflected continuum to give the total reprocessed
spectrum.

This total reprocessed spectrum can then be corrected for relativistic and
kinematic effects which cause ``smearing'' of spectral features (e.g.\
Fabian et al.\ 1989). These  effects are calculated for a monochromatic line
in the {\sc XSPEC} {\tt diskline} model so we use this, modified to include the
light bending effect in Schwarzschild geometry, to calculate the
spectral smearing, and convolve this with the total reprocessed spectrum.
The {\tt diskline} model is parameterised by the inner and outer radii of the
reflecting accretion disc, $\Rin$ and $\Rout$,  radial distribution
of irradiation emissivity for which we assume $\Firr(r)\propto r^{\beta}$
and the inclination of the disc, $i$.
In our basic models we assume $\Rout=10^4\,\Rg$ and
$\beta=-3$ (see Appendix A)
and fit $\Rin$ and the amplitude of the reprocessed component
expressed as the solid angle of the reprocessor as seen from the X--ray
source normalized to $2\pi$ i.e.\ $\Or\equiv\Omega/2\pi$ (so that $\Or=1$
corresponds to an isotropic source located above a flat, infinite disc).
This means that we do not assume any specific geometry which would result
in a unique relation between $\Rin$, $\Rout$, $\beta$ and $\Or$. The models
will be referred to as \relrepr hereafter. 
 
We assume cosmic abundances of Morrison \& McCammon (1983) with the
possibility of a variable iron abundance in the reprocessor; we fix the
inclination of the source at $i=56^{\circ}$ and assume the mass of the black
hole is $12\,\MSun$ (Shabaz et al.\ 1994).
 
The spectral analysis was performed using {\sc XSPEC} ver.\ 10 (Arnaud 1996)
into which all the non-standard models mentioned above were implemented as
local models.

\section{Results}

\begin{figure*}
 \epsfxsize = 0.95\textwidth
 \epsfbox[18 170 560 700]{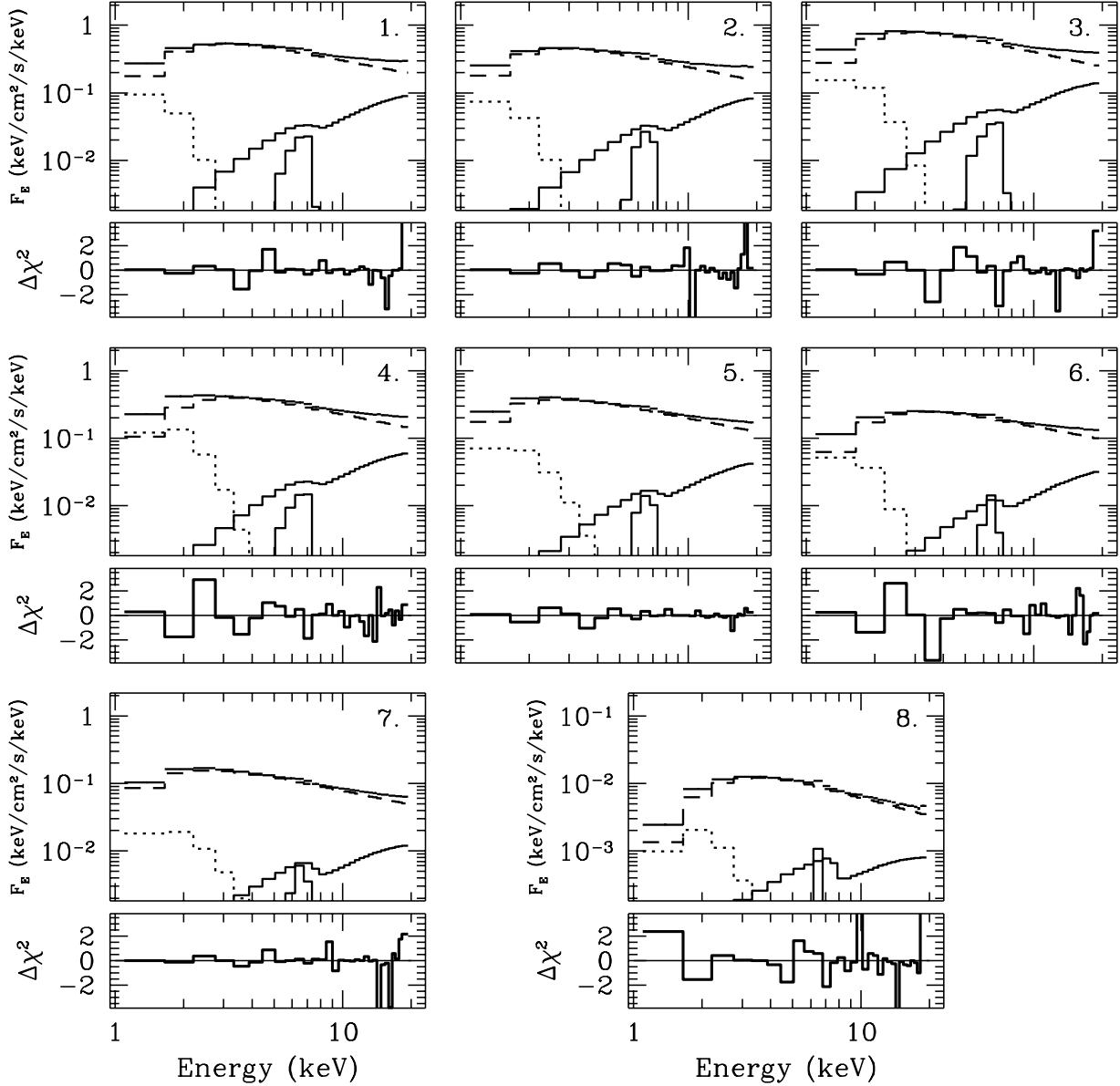}
 \caption{
 Unfolded spectra of the source (upper box in each panel) and $\chi^2$
 residuals (lower boxes) for all data sets. Labels in upper right corners refer
 to given data set (Table~\ref{tab:declspec}). Data points are plotted
 with (invisible) error bars. The solid histograms show the reprocessed
 component; the Fe $\Ka$ line and the Compton-reflected continuum are 
 separated for plotting only. The dashed histograms are the primary power 
 laws while
 the dotted histograms show the soft thermal component. Note that the
 flux scale in the last panel (8.) is shifted down by factor of 10.
 \label{fig:bigspec}}
\end{figure*}

The primary spectrum of the source is consistent with a simple power law after
1.~June (MJD-47000 = 677), so we can use this, together with the \relrepr model
and a soft component (disc blackbody) to describe the data.  We assumed our
basic reprocessing model, i.e.\ only one reprocessed component 
(cf.\ Section~\ref{sec:addrepr}), with ionization and relativistic smearing
allowed for. We fix $\Rout$ at $10^4\,\Rg$ and
$\Firr(r)\propto r^{-3}$ and we fit the amplitude $\Or$, the inner radius
$\Rin$ and the ionization parameter $\xi$. Results of the
spectral modelling are presented in Table~\ref{tab:declspec} and the spectra
are plotted in Figure~\ref{fig:bigspec}.

\begin{figure}
 \epsfxsize = 7.5 cm
 \epsfbox[0 220 512 680]{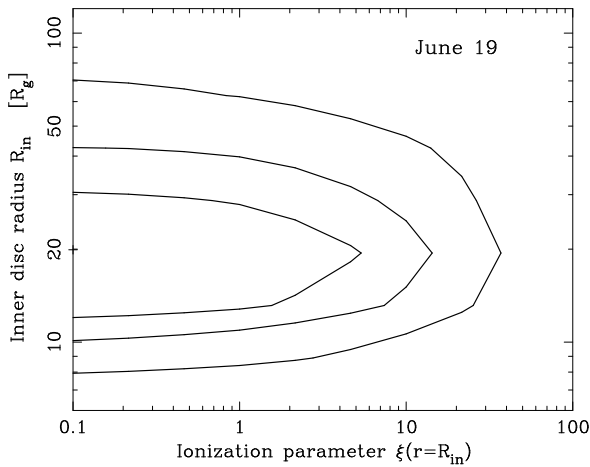}
 \caption{This plot demonstrates that the effects of relativistic smearing 
 and broadening due to
 multiple ionization {\it can\/} be separated in our modelling.
 It shows contours of $\Delta\chi^2 = 2.3,\ 4.6\ {\rm and}\ 9.2$ 
 as function of the inner disc radius $\Rin$ and the 
 ionization  parameter $\xi(r=\Rin)$ (assuming $\xi(r)\propto r^{-4}$)
 for the June 19.\ data set.
 The best fit has low ionization ($\xi\sim 0.1$) and $\Rin\sim 20$, and
 there is no correlation between the two parameters.
 \label{fig:rin_vs_xi0}}
\end{figure}

The reprocessed component is significantly present in all data sets.
It is very weakly ionized, and generally significantly smeared although not to
the extent expected for a disc extending down to the last stable orbit at 
$6\,\Rg$. We note that the two effects (ionization and relativistic smearing) 
can be separated given good quality data. The basic reason for this is that
the ionization broadens the spectral features towards higher energy whilst
the relativistic effects smear them around their rest--frame energy. Secondly,
the strength of the Fe spectral features increases dramatically with 
ionization whilst relativistic effects only redistribute the photons.
Figure~\ref{fig:rin_vs_xi0} illustrates the above discussion showing that
there is no correlation between $\xi$ and $\Rin$, even if a radial 
distribution of ionization is assumed.

\begin{figure*}
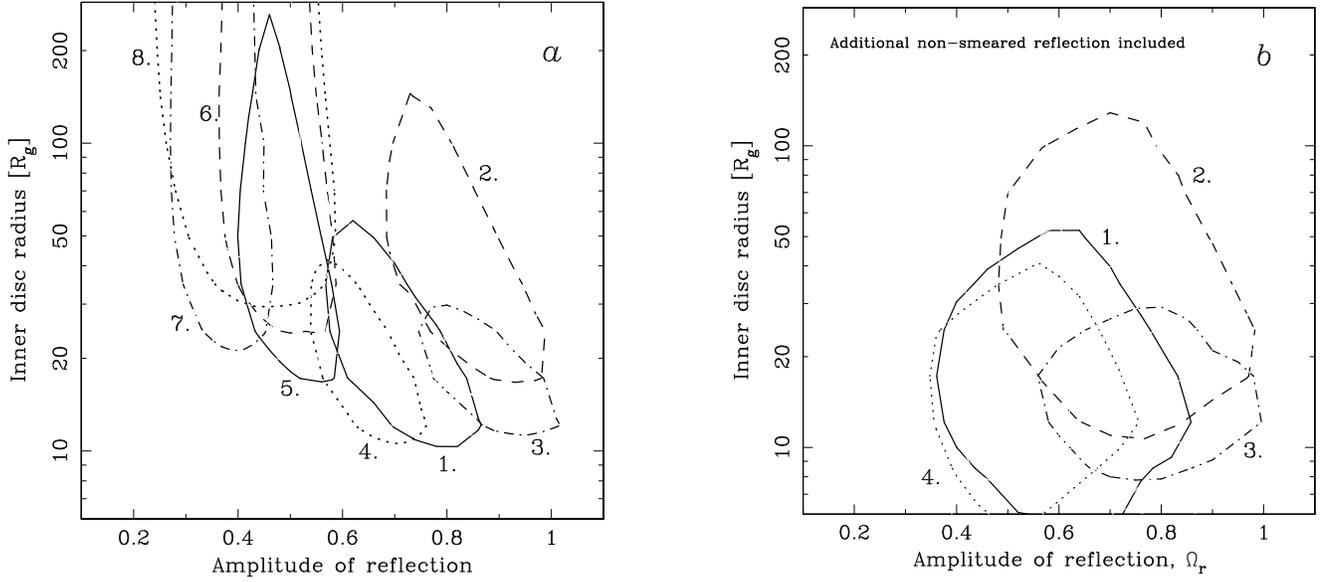

\parbox{\textwidth}{
 \parbox{0.48\textwidth}{
  \epsfxsize = 7.5 cm
  \epsfbox[10 150 500 680]{rin_vs_f.ps}
 }\hfill \parbox{0.48\textwidth}{
  \epsfxsize = 7.5 cm
  \epsfbox[50 150 550 680]{fe_refl_feschw_rin_vs_f.ps}
}}
 \caption{Contours of $\Delta\chi^2 = 4.61$ (90 per cent confidence limits
 for two parameters) as function of the amplitude of the smeared reprocessed
 component, $\Or$,
 and the inner disc radius, $\Rin$, derived from the relativistic smearing.
 Labels indicate different data sets (Table~\ref{tab:declspec}), from
 the earliest (June 3.) to the last observation (November). Panel {\it a\/}
 shows the contours when only the disc-reflected, smeared reprocessed 
 component
 is present, while panel {\it b\/} shows the same contours when an additional, 
 non-smeared reprocessing is allowed for in the model. The amplitude of that
 non-smeared component is constrained to be $<0.2$. Contours for only 
 the first  four data sets are plotted in {\it b\/}. Presence of the
 non-smeared component broadens the acceptable range of parameters of 
 the smeared, disc reflection.
\label{fig:declcont1}}
\end{figure*}

Figure~\ref{fig:declcont1}a shows how the amplitude of the reflected component
decreases during the decline phase. Initially, the amplitude of reflection 
is close to unity i.e.\ the cold, optically thick disc covers $\sim 2\pi$ 
solid angle as
seen from the X--ray source, but this drops significantly as the source
luminosity fades. The inner disc radius derived from the amount of relativistic
smearing is always incompatible with the last stable orbit in a Schwarzschild 
metric of
$\Rms=6\,\Rg$ (although with the caveat that $\Rin$ is correlated with the form
of irradiation emissivity). The inner disc radius is consistent
with being anti-correlated with the amount of reflection, although the 
change in the amount of smearing is only marginally significant. 

These results agree with and extend those
obtained in Paper I, where only a subset of present data was analyzed.
They confirm that optically thick accretion material (the putative accretion 
disc) was present in the system throughout the decline,
even though the source luminosity is clearly
dominated by the power law component, rather than by direct thermal emission
from the disc.

\section{COMPARISON WITH OTHER SXT}

\begin{figure*}
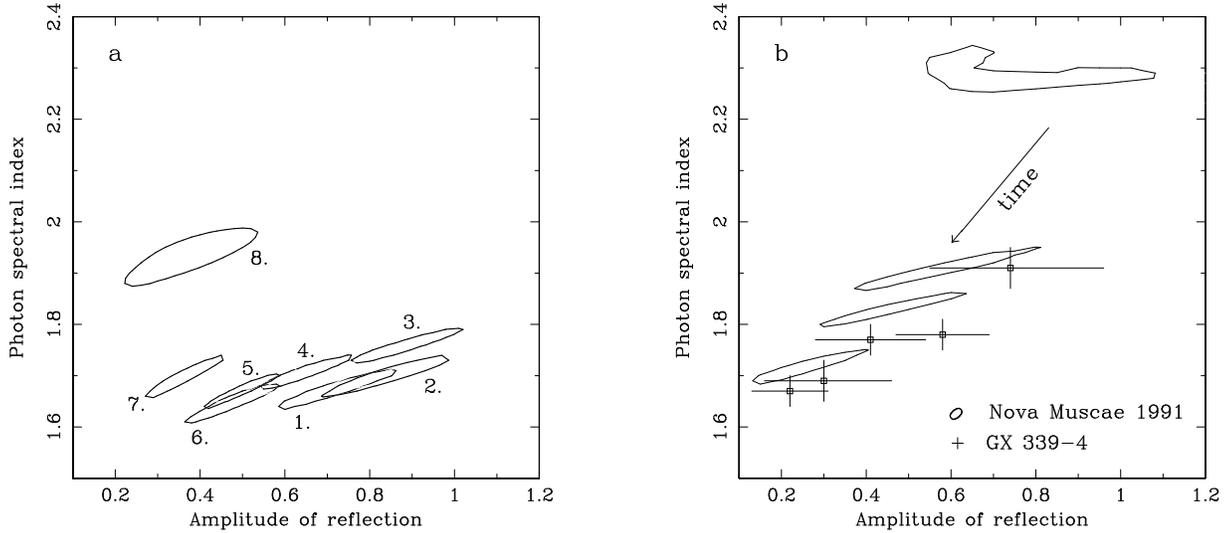

\parbox{\textwidth}{
\parbox{0.48\textwidth}{
 \epsfxsize = 7 cm 
 \epsfbox[10 150 522 650]{g_vs_f.ps}
}\hfil\parbox{0.48\textwidth}{
 \epsfxsize = 7 cm 
 \epsfbox[10 150 522 650]{gsx_g_vs_f.ps}
}}
 \caption{Contours of $\Delta\chi^2 = 4.61$ (90 per cent confidence limits
 for two parameters) as function of the amplitude of reflection, $\Or$,
 and the photon spectral index, $\Gamma$, for
 the decline phase spectra of GS~2023+338 (panel {\it a\/}; see 
 Table~\ref{tab:declspec}) and (panel {\it b\/}) for the soft and hard 
 states of Nova Muscae 1991 (ZDS98) and GX~339-4 (Ueda et al.\ 1994).
 The time arrow refers to Nova Muscae only.
\label{fig:declcont2}}
\end{figure*}

The trend we see in the data of a gradual decrease of $\Or$ possibly correlated
with increasing $\Rin$ is similar to that previously found in another SXT
source, Nova Muscae 1991 (ZDS98, Fig.~2).
However, there is a marked difference in the {\it spectral\/} evolution of
GS~2023+338 compared to Nova Muscae. The spectral index of the primary spectrum
of GS~2023+338 stayed roughly constant during the decline phase, except for the
last data set (November) obtained 5 months after the outburst, when the power 
law was significantly softer than during the first two months. Conversely in 
Nova Muscae we found a monotonic decrease of $\Gamma$ (hardening of the 
spectrum) with time, correlated with decreasing $\Or$ and increasing 
$\Rin$ (see Figure~\ref{fig:declcont2}; ZDS98). A similar correlation between  
$\Or$ and $\Gamma$ is also seen from GX339-4 (Ueda et al.\ 1994, 
Figure~\ref{fig:declcont2})

The results from Nova Muscae and GX339-4 can be understood in terms of a
phenomenological scenario where a retreating disc (increasing $\Rin$) gives a
smaller solid angle for reflection and a diminishing supply of soft photons for
Comptonization, so leading to a harder X--ray continuum spectrum (Esin et al.\ 
1997; ZDS98).  The lack of such a correlation in GS~2023+338 is then a clear
problem, implying that the observed change in solid angle for the reflecting
material does not give rise to a change in the ratio of soft seed photons
available for Compton scattering i.e.\ that the observed reflecting material is
not the dominant source of soft seed photons for the Comptonization which gives
rise to the hard power law. It is rather hard to imagine how this can be the
case. If there is another soft seed photon contribution, perhaps from stronger
cyclo/synchrotron emission, then the expected behaviour would be that the hard
X--ray source would have a softer spectral index. GS~2023+338 instead shows a 
rather hard spectral index at all times. Perhaps a better explanation is that 
there is an additional contribution to the reflected component from more 
distant, cold material, which does not then contribute much to the soft seed 
photons impinging on the X--ray source. We will now investigate such additional
reprocessed components.

\subsection{Additional sites of reprocessing}

\label{sec:addrepr}

\begin{figure}
  \epsfxsize = 7.5 cm
  \epsfbox[10 170 500 650]{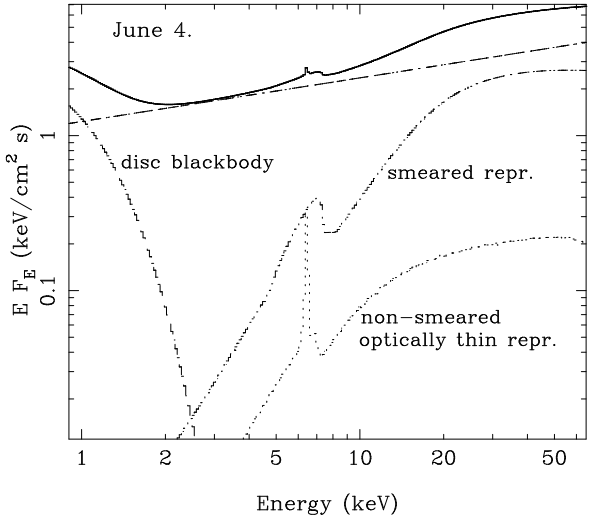}
 \caption{Model spectrum containing an additional, optically thin reprocessed
 component as fit to the June 4.\ data set. The amplitude of that reprocessed
 component is plotted at its 90 per cent upper limit (the best fit value is 
 factor of 10 lower). Such a component, due to reprocessing by cold plasma with
 $\NH \sim 4 \times 10^{23}\ {\rm cm^{-2}} $ i.e.\ $\tes \sim 0.3$, could be 
 present in the data until June 19. The evidence for presence of such plasma
 comes from rapid spectral variability which can be modelled 
 as photo--electric absorption (see Figure~\ref{fig:colcol}). Suppression
 of the Compton reflection bump due to $\tes<1$ is clearly seen in the
 model spectrum. Because of that, the required amplitude of the usual,
 optically thick reprocessing (possibly relativistically smeared) is still 
 high.
 \label{fig:thinreflspec}}
\end{figure}

Additional material around the source is clearly present, as evidenced by the
rapid spectral variability in the June 3.\ -- June 19.\ data (see
Figure~\ref{fig:colcol}) which can be attributed to photoelectric absorption
(Oosterbroek et al.\ 1997).  The reprocessed spectrum coming from this
(presumably distant) material would not be relativistically smeared and its
presence in the spectrum could mean smaller amplitude of the smeared, disc
reflected component. However, it is also probably not optically thick to
electron scattering. The spectra extracted from the extreme right (most
absorbed) end of the 'Nh track' in Figure~\ref{fig:colcol} give $\NH \approx
4\times 10^{23}\ {\rm cm^{-2}}$ (see also Oosterbroek et al.\ 1997). 
The Compton reflection hump from such reprocessor is therefore suppressed 
but the iron $\Ka$
line is not, since the optical depth for photo--electric absorption at the Fe
K-edge energy is $\sim 1$.  We have used our Monte Carlo code to simulate
reprocessing by such material including the iron line, assuming the reprocessor
is not ionized.  We then created a table model for {\sc XSPEC} (parameterised 
by $\Gamma$) and repeated our spectral fitting of the smeared reprocessing
including also this additional non-smeared component.
The best fit contribution of this component is rather small, with $\On\la
0.01$, but with 90 per cent upper limits ($\Delta\chi^2=2.7$) of $\On=0.19,\
0.22$ and 0.19, for June 3., June 4.\ and June 10.\ data sets, respectively.
However, this does not substantially change or broaden the allowed ranges of
$\Or$ of the disc-reflected, smeared component.  Figure~\ref{fig:thinreflspec}
shows the model fitted to the June 4.\ data set, with the contribution from the
non-smeared reprocessing at its 90 per cent upper limit, and it shows clearly
how the lack of a fully developed Compton hump in the optically thin reflection
component cannot contribute much to the observed spectral hardening beyond 10
keV, so as much optically thick reflection is needed as before.

Clearly, the observed (optically thin) absorbing material is insufficient to
produce an excess reflected component.  We cannot however exclude a priori a
possibility of an additional, optically thick reprocessed component being
present in the data.  Such component could come from e.g.\ an outer rim of the
accretion disc, since the SS solution predicts that the disc flares in its 
outer regions.

Solving the SS equations for vertically integrated, stationary, 
non-irradiated  disc  we obtain $\On\sim H(R=\Rout)/\Rout \la 0.03$ at
$\Rout=R_{\rm tidal}=8\times 10^{5}\ \Rg$ (for $\alpha=0.01$--0.5,
$\mdot=0.01$--0.05). Solutions with explicit vertical structure computed
by Meyer \& Meyer--Hofmeister (1982) give $H/R\,(\Rout) \sim 0.1$.
An irradiated disc would be expected to be more puffed up, but 
solutions of the vertical structure of irradiated discs 
depend on details of treatment of the X--ray absorption. Meyer \& 
Meyer--Hofmeister (1982) assumed that all incident X--rays are absorbed
and obtained significant thickening of a disc, up to $H/R\,(\Rout) \sim 0.3$.
On the other hand, observational estimates  of the optical (disc reprocessed) 
emission in Low  Mass X--Ray Binaries  also seem to
imply $H/R\sim 0.2$ (de Jong, van Paradijs \& Augusteijn 1996). 

%Using this as an upper limit to the 
%possible solid angle subtended by the outer disc gives the contours in
%the $\Or$ vs.\ $\Rin$ plane plotted in Figure~\ref{fig:declcont1}b, which are
%then rather broader than the original ones where all the 
%reflection was due to the relativistic disc. 

We therefore added a cold ($\xi=0$), non-smeared reprocessed
component (continuum with iron $\Ka$ line) to our basic model and we repeated
fits to first four data sets. In view of the above discussion we set an 
upper limit of 0.2 to its amplitude. We found that a substantial contribution 
from such a component is indeed allowed by the data. Consequently, the
contribution of the smeared component can be smaller. 
The results in the $\Or$--$\Rin$ plane for the {\it relativistic\/}
reprocessing now allow for almost constant $\Rin$ and $\Or$
(Figure~\ref{fig:declcont1}b).

\begin{figure}
  \epsfxsize = 7.5 cm
  \epsfbox[10 180 550 700]{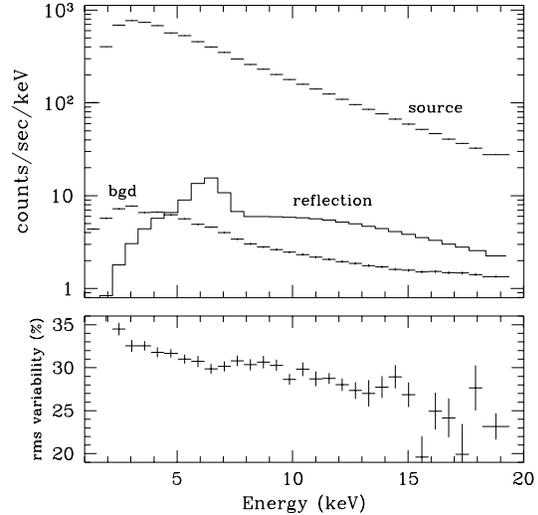}
 \caption{Narrow spectral features in model spectra (e.g.\ an emission lines) 
  are broadened by detector
  response, so their presence does not manifests as a channel-to-channel 
  change of e.g.\ rms variability in the observed spectrum.
  The upper panel shows the count spectrum (i.e.\ as registered by \ginga LAC)
 of the source on June 10 (data set 3.) showing the source, background and
 possible contribution from the model {\em non-smeared\/} reprocessed 
 component  (at it 90 per  cent upper limit, 
 cf.\ Figure~\ref{fig:thinreflspec}).  Lower panel shows the 
 observed rms variability (computed from  non-bgd-subtracted data). 
 \label{fig:j10_rms}}
\end{figure}

Any reprocessed component coming from the distant outer rim of the disc
should be constant or only slowly variable. Given that the
reflected spectrum is hard, then this should reduce the amount of variability
seen at high energies, and around the narrow iron line. Oosterbroek et al.\
(1997) show that there is no obvious effect at the iron line energy for any of 
the decline spectra (although there is a marked constant iron line component 
during the outburst). However, this does not rule out the amount of unsmeared
reprocessing required here. Figure~\ref{fig:j10_rms} shows the June 10th data 
set, together with the maximum contribution from the non-smeared reflection. 
Even though the line is intrinsically narrow, it is broadened by the response
of the detector. At the iron line energy the narrow component contributes 
only 3 per cent of the counts, so its effect on the variability is
small. Figure~\ref{fig:j10_rms} also shows the observed 0.01--8 Hz rms 
variability spectrum of the June 10 data, showing that the variability pattern
cannot rule out possible presence of an unsmeared, constant reflected 
component with $\On\sim 0.2$.

We are left then with the far from satisfactory situation where GS~2023+338
shows a very different behaviour to that of other SXT. Either it has a
remarkable outer disc, which subtends a surprisingly large solid angle, or the
source is somehow able to maintain a constant hard X--ray spectral shape 
despite there being a copious and changing amount of soft photons from the 
nearby accretion disc.

\section{DISCUSSION}

The results from the amplitude of reflection and amount of relativistic 
smearing imply that the disc does not extend down to the last stable orbit 
at $6\, \Rg$. 
An independent estimate of the inner disc radius could in principle be obtained
from the amplitude of the thermal emission, which is proportional to
the emitting area. The amplitude is not always well constrained for our data
because
only the Wien cutoff is within the \ginga LAC band--pass for the observed 
temperatures.
In those cases where the amplitude is well constrained the inner radius
can be computed from
\[
\Din = D_{10}\,\sqrt{{K \over \cos i}}\,\left({\Tcol \over \Teff}\right)^2
\ \ \ \ {\rm [km]},
\]
where $K$ is the normalization of the {\tt diskbb} model in 
{\sc XSPEC}, 
$D_{10}$ is the distance to the source in units of 10 kpc, $i$ is the
inclination angle and $\Tcol/\Teff$ is the ratio of colour to effective
temperatures, usually in the range of 1.5--1.9 (e.g.\ Shimura \& Takahara 
1995).

The inferred inner disc radius ($\Din$ in Table~\ref{tab:declspec} -- N.B.
values shown are computed without the colour temperature correction) 
is compatible with that inferred from relativistic smearing until June 19th.
$\Din$ is smaller than $\Rin$ afterwards, although a presence of a disk
blackbody component of amplitude corresponding to $\Din=\Rin$ cannot be 
excluded, if its temperature was below $\sim 0.18$ keV, again due to limited
band--pass of the LAC detector.

The fact that $\Din$ and $\Rin$ are in agreement is not easily compatible
with strong constraints on ionization of the reprocessor.
Throughout the entire decline phase the reflecting material had to
be very weakly ionized, with typical upper limit on the mean ionization
stage of iron $<$Fe$\,${\sc V}. Even assuming a strong radial distribution
of ionization, $\xi(r) = \xi(\Rin)\,(r/\Rin)^{-4}$, and allowing for the
additional non-smeared and cold reprocessing (Section~\ref{sec:addrepr}), 
we obtain the highest allowed
Fe ionic stage of Fe$\,${\sc XVI}. This corresponds to the reprocessor
temperature $<(2-4)\times 10^{5}\,{\rm K} \approx 20-40\,{\rm eV}$, 
largely independently of whether iron ionization is in LTE conditions 
(e.g.\ Rybicki \& Lightman 1979) or  photoionization dominates 
(Figure 2 in \.{Z}ycki \& Czerny 1994), and significantly smaller than
the temperature of the observed soft emission.

The observed (weakly significant) increase of $\xi$ in July and November
data is not consistent with a stationary SS disc, since the latter 
solution predicts a weaker  $\mdot$ dependence of the density
$n(\mdot) \propto \mdot^{11/20}$ than $\LX(\mdot)$ which presumably is
$ \propto \mdot$,  hence $\xi(\mdot)\propto \mdot^{9/20}$ should
decrease as $\mdot$ decreases. Similarly, radial dependence of the density,
$\propto r^{-15/8}$ is slower than that of the irradiating flux, 
$\propto r^{-3}$. This may mean that the disc is not in a steady
state; instead the disc may have returned to cold, low viscosity state in
its outer region where a mass build-up has begun, cutting off the supply of
mass to the inner disc and causing departure from stationary solution for
the inner disc.
However, solving the vertically averaged SS disc equations (for an 
un-irradiated disc) we find that Hydrogen recombines only beyond 
$r\sim 10^{5}\,\Rg$ when $\mdot\sim 10^{-3}\,\MEdd$, and so the stationary
disc solution should be valid at the much closer radii where the reflection
comes from.

The power law spectral index in the last data set (November, 155 days
after outburst) is  significantly steeper than earlier 
($\Gamma\sim 1.9$ vs.\ $\Gamma\sim 1.7$; Table~\ref{tab:declspec}). 
Since the source luminosity at that time 
was $\sim 3\times 10^{-4}\,\LEdd$ it may seem the source was close to
quiescence and consequently its emission was dominated by (Comptonized)
synchrotron radiation, as in ADAF-based model of Narayan, Barret \& 
McClintock (1997), and
comparable to what was observed by ASCA. However, we do see the reprocessed
component with $\Or\sim 0.3$ so, consequently, the emission should be 
dominated by comptonization of the soft photons due to thermalized irradiation,
if the reflection was indeed from the inner disc. Alternatively, again 
a significant
contribution from outer disc is a possibility, with no feedback of soft
photons to the X--ray source. However, the required magnitude of disc flaring
is again rather larger than expected. We also note that if the outer disc
contribution to reprocessing is dominant at the end of the decline phase,
then the overall decrease of $\Or$ has to be attributed to decrease of
the smeared component, again rising the problem of the unchanging spectral
shape.

\section{CONCLUSIONS}

Our results support growing evidence that the geometry of accretion 
flow evolves as the accretion rate changes during the decline of SXT.
The evolution involves a retreat of the inner, optically thick disc
as indicated by both decreasing amplitude of the reprocessed component and 
the relativistic smearing becoming  less or insignificant (Paper I; ZDS98).
These variations of geometry, observed also in persistent source Cyg X-1
(Done \& \.{Z}ycki 1998) are usually correlated with spectral states: small
inner disc radius corresponds to soft spectra whilst large radius corresponds
to hard spectra (Nova Muscae 1991, ZDS98). 
This correlation is easily understood in terms of a scenario
where the availability of soft seed photons for Comptonization, determined
by the inner disc radius, regulates the Compton cooling and, consequently,
the spectral shape (Esin et al.\ 1997).

However, while \gs showed the same receding inner disc, its hard X--ray
continuum spectral shape remained largely constant.
This poses serious problems to any model
involving Comptonization as a source of hard X--rays, and in which changes 
in the spectral properties  are linked to the geometry  of accretion.

\section*{Acknowledgements}

CD acknowledges support from a PPARC Advanced Fellowship.
Work of PTZ was partly supported by grant no.\ 2P03D00410 of the Polish 
State Committee for Scientific Research. 
%We thank the anonymous referee
%for pointing out the importance of the colour temperature correction.

\appendix

\section{Radial dependence of irradiation emissivity}

\begin{figure*}
 \epsfxsize = 0.9\textwidth
 \epsfbox{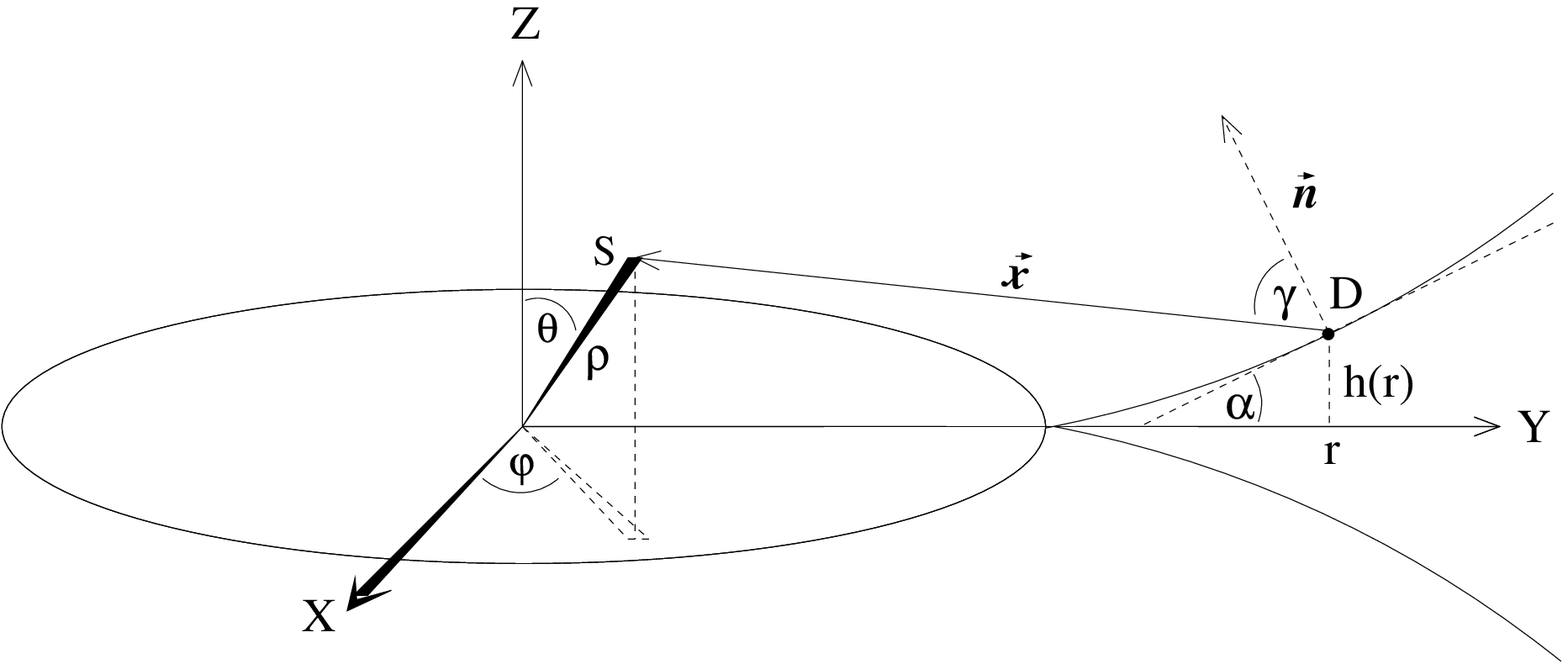}
 \caption{Geometry of disc irradiation. The X--ray source is assumed spherical,
 with outer radius $R_0$. The disc extends from $R_0$ to $\Rout\gg R_0$.
 S is an arbitrary point within the source while D is a point on the disc 
 surface.  Note that the X--ray source is 
 plotted in perspective  but a cut in y-z plane for the disc is used.
\label{fig:irrad_geom}}
\end{figure*}

The irradiation emissivity follows closely the $r^{-3}$ 
dependence if the reflecting disc is outside a central source of radiation,
except very close to the source.
Consider a spherical source, centered on the black hole, of maximum radius
$R_0$, with the disc extending outside it, i.e.\ from $r=R_0$ to 
$\Rout\gg R_0$. We allow for a flaring disc with height described by an 
arbitrary
function  $h(r)$. Introducing spherical coordinate system in the usual manner
(see Figure~\ref{fig:irrad_geom}) we have:
\[
{\rm S} = [\rho\, \sin\theta\, \cos\phi, \rho\, \sin\theta\, \sin\phi,
\rho\, \cos\theta],
\]
\[
{\rm D} = [0, r, h(r)].
\]
Vector normal to the disc surface is
\[
\bmath{n} = \left[0,-{d h\over d r},1 \right].
\]
Distance between S and D is the length of the vector from S to D,
$l\equiv |\bmath{x}|$, 
\begin{equation}
 l^2 = \rho^2+r^2 - 2\rho r\sin\theta \sin\phi +h(r)[h(r)- 2\rho \cos\theta],
\end{equation}
while the cosine of the illumination angle is
\begin{eqnarray}
\lefteqn{\cos\gamma = \bmath{n}\cdot\bmath{x}/(|\bmath{n}| |\bmath{x}|) = } 
 \nonumber \\
 & & { {d h\over d r}(r-\rho\sin\theta \sin\phi) + \rho\cos\theta-
 h(r) \over l\,\sqrt{1+(d h/dr)^2}} . 
\end{eqnarray}
Contribution to the irradiation emissivity is thus
\begin{equation}
\delta\Firr = d V \, \cos\gamma/l^2,
\end{equation}
where $d V=\rho^2 \,\sin\theta\, d\rho\,d\theta\,d\phi$ is the volume element
of the source.
Neglecting radiative transfer and relativistic effects within the source,
we obtain the emissivity by integrating $\delta\Firr$ over the source,
\[
\Firr(r) = \int f(\bmath{r})\, \delta\Firr,
\]
where $f(\bmath{r})$ describes the distribution of volume emissivity
within the source.

\begin{figure}
\epsfxsize = 0.45\textwidth
\epsfbox[18 250 592 718]{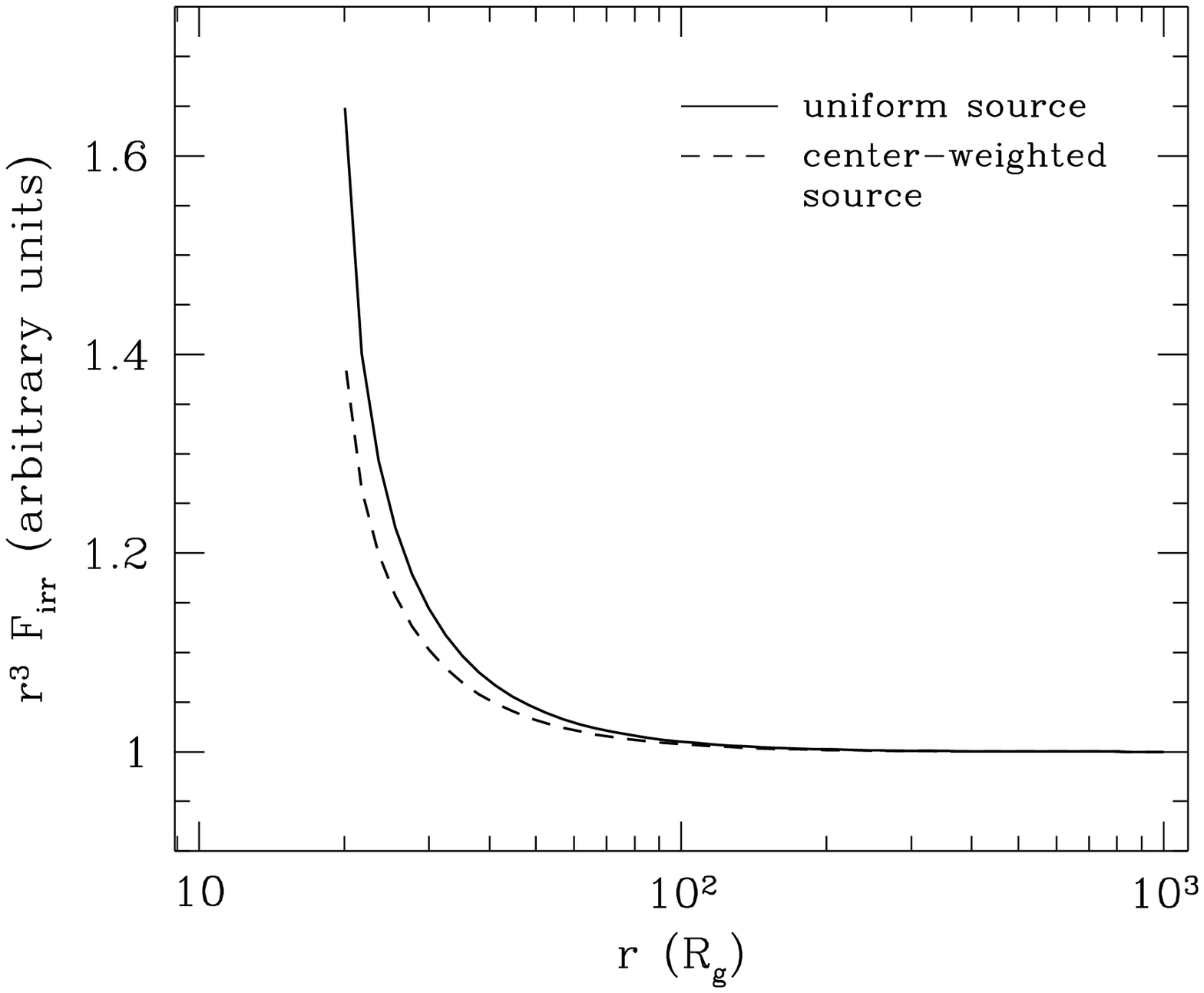}
\caption{Irradiation emissivity, $r^{3} \Firr(r)$, for a central, spherical 
 source illuminating an external, flat disc. Inner source radius is assumed
 $6 \Rg$, outer one $20 \Rg$, from where  the disc extends outwards. 
 The emissivity
 is somewhat steeper than $r^{-3}$ for $r<100 \Rg$ and its shape depends
 on how strongly the luminosity inside the source is concentrated towards
 the centre. The solid curve shows the emissivity for a uniform source, whilst
 the dashed curve is for  the luminosity per unit volume 
 $r^{-4} (1-\sqrt{6/r})$.
 In the $20 - 40 \Rg$ range the approximate power law fit to $\Firr(r)$
 has slope equal $\beta=3.5$ in the former case, while $\beta=3.3$ in the 
 latter.
\label{fig:firr}}
\end{figure}

Examples of $\Firr(r)$ are plotted in Figure~\ref{fig:firr}. As can be seen
close to the source the emissivity can deviate from the $r^{-3}$ dependence.
It is generally steeper ($r^{-\beta}$ with $\beta>3$) due to the finite size
of the source. The magnitude of deviation depends on luminosity distribution 
within the source. The stronger the central concentration of luminosity, the
weaker the deviations.

A special case of a point source located on the symmetry axis, above a disc
is recovered from (A1) and (A2) assuming polar angle $\theta=0$. 
Assuming further a flat disc ($h(r)=0$) we obtain well known formula,
\[
\Firr(r) \propto {1 \over \left(r^2 + h_{\rm s}^2\right)^{3/2}} \simeq
r^{-3}\ {\rm for}\ r\gg h_{\rm s},
\]
where $h_{\rm s}$ is the height of the source above the disc plane.

Another frequently quoted formula for the irradiation 
temperature in a flaring disc (SS, van Paradijs 1996, King et al.\ 1997), 
results from our formulae (A1) and (A2) in the limiting case of $\rho=0$,
putting $l = r$ (i.e.\ it is equivalent to assuming $r\gg R_0$) and 
neglecting $d h/d r$ in denominator of (A2).

Our formulae (A1)--(A3) could also in principle be applied to a disc extending 
{\em inside\/}  the source. A simpler way to find the irradiation emissivity 
in this case is to notice
that $\Firr(r)$ follows the local electron emission, vertically integrated
over the hot source, $q^{-}$. This generally follows the viscous energy 
generation in the hot source which however is not necessarily described
by the usual dependence  $q^{+} \propto  r^{-3}$.
First, the ratio of $q^{+}_{\rm hot}/q^{+}_{\rm tot}$ can be a function of 
radius (Witt, Czerny \& \.{Z}ycki 1997). Second, a fraction $f$ of 
$q^{+}_{\rm hot}$ can be advected, and $f$ itself can be a function of radius.
Third, Compton cooling by non-local soft photons can have a strong 
effect (Esin 1997), especially in the transition region. Since the non-local 
cooling enhances electron emission (Esin 1997), we can expect $\Firr(r)$
to be steeper than $r^{-3}$ i.e.\ $\beta>3$ close to the inner disc radius.
All the above effects are model--dependent and need to be properly modelled.

{}

\end{document}